\documentclass{article}

\usepackage{arxiv}
\usepackage[utf8]{inputenc} % allow utf-8 input
\usepackage[T1]{fontenc}    % use 8-bit T1 fonts
\usepackage{hyperref}       % hyperlinks
\usepackage{url}            % simple URL typesetting
\usepackage{booktabs}       % professional-quality tables
\usepackage{amsfonts}       % blackboard math symbols
\usepackage{nicefrac}       % compact symbols for 1/2, etc.
\usepackage{microtype}      % microtypography
\usepackage{lipsum}		% Can be removed after putting your text content
\usepackage{graphicx}
\usepackage[super,sort&compress,comma]{natbib}
\usepackage{doi}
\usepackage{chemformula}
\usepackage[nohyperlinks, nolist]{acronym}
\usepackage{subcaption}
\usepackage{physics}
\newcommand{\orcidlink}{https://orcid.org/}

\newcommand{\lucaorcid}{0000-0003-0807-682X}
\newcommand{\lauraorcid}{0000-0002-9760-5465}
\newcommand{\peterorcid}{0000-0003-1611-3395}
\newcommand{\mattorcid}{0009-0005-4426-4571}
\newcommand{\niklasorcid}{0009-0000-0237-2543}
\makeatletter
\renewcommand{\@cite}[1]{
{$\!\! ^{(#1)}$}}
\makeatother
\newcommand{\scaling}{\varphi}
\newcommand{\wavelet}{\psi}

\title{Combining the Maximum Overlap Method with Multiwavelets for Core-Ionisation Energy Calculations}

\date{\today}	% Here you can change the date presented in the paper title
\author{ \href{\orcidlink\niklasorcid}{\includegraphics[scale=0.06]{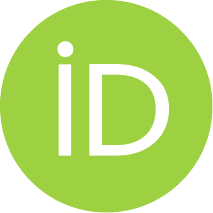}\hspace{1mm}}Niklas G\"{o}llmann\\
Theoretische Organische Chemie,\\ Organisch-Chemisches Institut and Center for Multiscale Theory and Computation,\\ 
Universit\"{a}t M\"{u}nster, Corrensstra{\ss}e 36, 48149 M\"{u}nster, Germany \thanks{Secondary address: Hylleraas Centre, Department of Chemistry, UiT The Arctic University of Norway, N-9037 Tromsø, Norway \\}\\
\And
\href{\orcidlink\mattorcid}{\includegraphics[scale=0.06]{orcid.pdf}\hspace{1mm}}Matthew~R.~Ludwig\\
Centre for Computational Chemistry,\\ School of Chemistry, University of Bristol,\\ Bristol BS8 1TS,  United Kingdom \\
\And
\href{\orcidlink\peterorcid}{\includegraphics[scale=0.06]{orcid.pdf}\hspace{1mm}}Peter Wind\\		%% Affiliation \\
Hylleraas Centre, Department of Chemistry,\\ UiT The Arctic University of Norway,\\ N-9037 Tromsø, Norway \\
\And
\href{\orcidlink\lauraorcid}{\includegraphics[scale=0.06]{orcid.pdf}\hspace{1mm}}Laura Ratcliff \\
Centre for Computational Chemistry,\\ School of Chemistry, University of Bristol,\\ Bristol BS8 1TS, United Kingdom $^{*}$\\ \texttt{laura.ratcliff@bristol.ac.uk}
\And
\href{\orcidlink\lucaorcid}{\includegraphics[scale=0.06]{orcid.pdf}\hspace{1mm}}Luca Frediani\\
Hylleraas Centre, Department of Chemistry,\\ UiT The Arctic University of Norway,\\ N-9037 Tromsø, Norway \\
}

\renewcommand{\headeright}{Preprint}
\renewcommand{\undertitle}{Preprint}
\renewcommand{\shorttitle}{Core-Ionisations with MOM and Multiwavelets}
\hypersetup{
pdftitle={Core-Ionisation with MOM and Multiwavelets},
pdfsubject={Quantum Chemistry},
pdfauthor={},
pdfkeywords={Multiwavelets, Wavelets, Real-space, MOM, X-RAY, core energy, binding energy},
}

\begin{document}
%
% This is a random collection of acronyms
% Acronyms are ordered alphabetically by abbreviation
% Usage
% 1. add the following lines to your main tex document
% In the preamble: \usepackage{acronym}
% In the document: \input{acronyms}
%
% 2. RTFM of the acronym package!
%
% 3. Add/modify/remove the listes acronyms according to your needs
%
\begin{acronym}
\acro{AO}{atomic orbital}
\acro{AE}{all electron}
\acro{API}{Application Programmer Interface}
\acro{AUS}{Advanced User Support}
\acro{BE}{Binding Energy}
\acrodefplural{BE}[BEs]{Binding Energies}
\acro{BEM}{Boundary Element Method}
\acro{BO}{Born-Oppenheimer}  
\acro{CBS}{complete basis set}
\acro{CC}{Coupled Cluster}
\acro{CTCC}{Centre for Theoretical and Computational Chemistry}
\acro{CoE}{Centre of Excellence}
\acro{DC}{dielectric continuum}  
\acro{DD}{domain decomposition}
\acro{DFT}{density functional theory}  
\acro{DKH}{Douglas-Kroll-Hess}
\acro{EFP}{effective fragment potential}
\acro{ECP}{effective core potential}
\acro{EU}{European Union}
\acro{FMM}{fast multipole method}
\acro{FCI}{Full Configuration Interaction}
\acro{GGA}{generalized gradient approximation}
\acro{GPE}{Generalized Poisson Equation}
\acro{GTO}{Gaussian Type Orbital}
\acro{HF}{Hartree-Fock}  
\acro{HPC}{high-performance computing}
\acro{Hylleraas}[HC]{Hylleraas Centre for Quantum Molecular Sciences}
\acro{IEF}{Integral Equation Formalism}
\acro{IEFPCM}{Integral Equation Formalism \ac{PCM}}
\acro{IGLO}{individual gauge for localized orbitals}
\acro{IMOM}{Initial Maximum Overlap Method}
\acro{KB}{kinetic balance}
\acro{KS}{Kohn-Sham}
\acro{LAO}{London atomic orbital}
\acro{LAPW}{linearized augmented plane wave}
\acro{LDA}{local density approximation}
\acro{MAD}{mean absolute deviation}
\acro{maxAD}{maximum absolute deviation}
\acro{MM}{molecular mechanics}  
\acro{MCSCF}{multiconfiguration self consistent field}
\acro{MPA}{multiphoton absorption}
\acro{MRA}{multiresolution analysis}
\acro{MSDD}{Minnesota Solvent Descriptor Database}
\acro{MOM}{Maximum Overlap Method}
\acro{MW}{multiwavelet}
\acro{NAO}{numerical atomic orbital}
\acro{NeIC}{nordic e-infrastructure collaboration}
\acro{KAIN}{Krylov-accelerated inexact Newton}
\acro{NMR}{nuclear magnetic resonance}
\acro{NP}{nanoparticle}  
\acro{OLED}{organic light emitting diode}
\acro{PAW}{projector augmented wave}
\acro{PBC}{Periodic Boundary Condition}
\acro{PCM}{polarizable continuum model}
\acro{PSP}{pseudopotential}
\acro{PW}{plane wave}
\acro{QC}{quantum chemistry}  
\acro{QM/MM}{quantum mechanics/molecular mechanics}  
\acro{QM}{quantum mechanics}  
\acro{RCN}{Research Council of Norway}
\acro{RDM}{1-body reduced density matrix}
\acro{RMSD}{root mean square deviation}
\acro{RKB}{restricted kinetic balance}
\acro{SC}{semiconductor}
\acro{SCF}{self-consistent field}
\acro{SCRF}{self-consistent reaction field}
\acro{STSM}{short-term scientific mission}
\acro{SAPT}{symmetry-adapted perturbation theory}
\acro{SERS}{surface-enhanced raman scattering}
\acro{STO}{Slater-Type Orbital}
\acro{WPREL}[WP1]{Work Package 1}
\acro{WPROP}[WP2]{Work Package 2}
\acro{WPAPP}[WP3]{Work Package 3}
\acro{WP}{Work Package}  
\acro{X2C}{exact two-component}
\acro{ZORA}{zero-order relativistic approximation}
\acro{HAXPES}{hard X-ray photoelectron spectroscopy}
\acro{XPS}{X-ray photoelectron spectroscopy}
\end{acronym}

\maketitle

\begin{abstract}
	We present a protocol for computing core-ionisation energies for molecules, which is essential for reproducing X-Ray photoelectron spectroscopy experiments. The electronic structure of both the ground state and the core-ionised states are computed using Multiwavelets and Density-Functional Theory, where the core ionisation energies are computed by virtue of the $\Delta$SCF method. To avoid the collapse of the core-hole state or its delocalisation, we make use of the Maximum Overlap Method, which provides a constraint on the orbital occupation. Combining Multiwavelets with the Maximum Overlap Method allows for the first time an all-electron calculation of core-ionisation energies with Multiwavelets, avoiding known issues connected to the use of Atomic Orbitals (slow convergence with respect to the basis set limit, numerical instabilities of core-hole states for large systems). We show that our results are consistent with previous Multiwavelet calculations which made use of pseudopotentials, and are generally more precise than corresponding Atomic Orbital calculations. We analyse the results in terms of precision compared to both Atomic Orbital calculations and Multiwavelets+pseudopotentials calculations. Moreover, we demonstrate how the protocol can be applied to target molecules of relatively large size. Both closed-shell and open-shell methods have been implemented.  
\end{abstract}

% keywords can be removed

\section{Introduction}

Core \ac{XPS} is a powerful technique for probing the electronic structure of materials, from molecules to surfaces to solids. By providing a direct measurement of core \acp{BE}, that is the energy required to remove a particular core electron from the material, it is able to probe the \emph{local} electronic structure, while offering insights into bonding nature and chemical and coordination environments. For valence XPS, there is a long-standing tradition of using theory, often based on density-functional theory (DFT)~\cite{Hohenberg1964,Kohn1965} calculations, to interpret experimental spectra. For core XPS, this has not traditionally been the case, with interpretation instead typically relying on comparisons to reference spectra. This is challenging, if not impossible, for core spectra of more complex materials, where there may be many overlapping peaks and no clear way to assign a given peak to a particular atom. This has motivated an increasing interest in methods for simulating core \acp{BE}, opening up the possibility of gaining new insights into complex materials using core XPS.

A range of approaches exist for calculating core \acp{BE} using DFT, including Koopmans', the $Z+1$ or equivalent core model~\cite{Jolly1970,Bagus2020} and $\Delta$SCF~\cite{Bagus1965}. Of these, $\Delta$SCF is by far the most popular, and has been successfully applied to an increasingly wide range of systems~\cite{Vines2018,besley_self-consistent-field_2009,hait_orbital_2021}. Due to the challenges involved with calculating absolute \acp{BE}, it is common practice to align theoretical results with respect to experimental spectra. In such cases, relative core \acp{BE} are therefore sufficient for aiding in peak assignments and thus interpreting experimental results. However, $\Delta$SCF has also shown promise for calculating absolute core \acp{BE}~\cite{PueyoBellafont2016,Kahk2019}. An alternative approach to DFT is \emph{GW}~\cite{Hedin1965}, which has in recent years also been applied to core \ac{BE} calculations, e.g.\ as in Refs~\cite{Golze2020,Mejia-Rodriguez2021,Li2022}. Through a set of benchmark calculations to a test set of 65 molecules, covering C, N, O and F 1$s$ and also including relativistic effects, $\Delta$SCF and \emph{GW} were shown to have similar accuracies with respect to experiment~\cite{Golze2020}.

Like most molecular DFT calculations, the majority of $\Delta$SCF calculations in gas phase have typically employed Gaussian basis sets. However, convergence with respect to basis set size is slow, with smaller basis sets typically not being flexible enough to treat core-excited states~\cite{Fouda2017}, so that there is poor error cancellation between the ground-state and core-hole calculations. Accurate core \ac{BE} calculations in Gaussian basis sets therefore usually require large basis sets, adding to the computational cost. Alternative strategies based on e.g.\ supplementing a basis set with additional functions in an approach akin to the $Z$+1 approximation have been shown to provide a significant increase in accuracy without requiring too many additional basis functions~\cite{Hanson2018}. However, a more general way to overcome the limitations of Gaussian basis sets and reach high precision is to use a systematic basis set. One such alternative that has gained a prominent role in the past few years is constituted by \acp{MW}\cite{Harrison.10.1063/1.1791051,Jensen.10.1021/acs.jpclett.7b00255,Yanai.10.1063/1.1768161,Jensen.10.1039/c6cp01294a,Brakestad.10.1021/acs.jctc.0c00128,Bischoff.10.1016/bs.aiq.2019.04.003,Vence.10.1103/physreva.85.033403}. Multiwavelets are a specific realisation of wavelets, which use a set of polynomials on an interval~\cite{Alpert.10.1137/0524016,Alpert.10.1006/jcph.2002.7160}. The basis can be refined systematically and adaptively, providing strict error control on energetics and molecular properties. The comparatively large memory footprint of such a method has prevented its widespread use in the past, but modern \ac{HPC} architectures combined with efficient implementation has lifted this limitation except for extremely large systems~\cite{Wind2023}. Moreover, several studies have shown that \acp{MW} become competitive in performance if high precision is requested~\cite{Wind2023, Pitteloud.10.1021/acs.jctc.3c00693, Gubler.10.1021/acs.jpca.4c06708}.

Another challenge of $\Delta$SCF calculations is that of converging the core hole. Without constraints, the core hole may either delocalise or `hop' between core orbitals of other atoms of the same species~\cite{Bagus1972,Tolbatov2017}, particularly when those orbitals have similar energies, e.g.\ due to having the same local chemical environment. This can lead to poor convergence, or indeed convergence to the wrong solution, and thus ultimately significant errors in core \acp{BE}. 
This combination of challenges motivated previous work in which the $\Delta$SCF approach was implemented in a \ac{MW} framework~\cite{Pi2020}. In this work, a combination of an \ac{AE} and \ac{PSP} based approach was used, wherein the core states of the core-excited atom were treated explicitly, with the remainder of the atoms treated at the PSP level. The elimination of all other core orbitals prohibits the possibility of core-hole hopping, while the automatic refinement of the \ac{MW} approach adapts to provide a more refined (accurate) grid where needed, thereby enabling accurate core \ac{BE} calculations. 

In this work, we combine \acp{MW} with an alternative approach to ensure the core hole remains sufficiently localised, namely the \ac{MOM}~\cite{gilbert_self-consistent_2008} method. MOM, as well as its variant \ac{IMOM}~\cite{barca2018}, aim to keep the core hole localised on the correct atom by maximising the overlap between orbitals. By combining \ac{MOM} and \acp{MW} for the first time, we demonstrate the possibility of achieving both high numerical precision and robust convergence for $\Delta$SCF-based core \ac{BE} calculations of molecules. We first provide an overview of the theory, including \acp{MW}, $\Delta$SCF, and \ac{MOM} and \ac{IMOM}. We then present results for small molecules, amino acids and a large molecule, including comparisons with both Gaussian basis sets and the combined \ac{AE}/\ac{PSP} \ac{MW} approach. Finally, we finish with a summary and conclusion.

\section{Theory}

\subsection{Multiwavelets}

Wavelets and \acp{MW} are a family of functions first constructed in the 1980s, and initially designed for signal processing\cite{keinert}. They are localised both in real and Fourier space\cite{Strang.10.1090/s0273-0979-1993-00390-2}, which makes them ideal for obtaining compact representations, which in turn lead to fast algorithms. It was about 20 years ago that their potential was first realised for electronic structure calculations\cite{Cho.10.1103/physrevlett.71.1808}, leading to several practical implementations. For \ac{AE} calculations, the first code to demonstrate their potential was MADNESS\cite{Harrison.10.1137/15m1026171}, followed by MRChem\cite{bast_mrchem_2024,Jensen.10.1021/acs.jpclett.7b00255} a few years later. To date, these are still the only two codes worldwide which are capable of \ac{AE} \ac{MW} electronic structure calculations for energies and molecular properties.
They are both based on Alpert's realization of \acp{MW}, which in practice define the mother scaling functions as a set of polynomials $\scaling_j{}, j=0,k$ up to order $k$ on an interval\cite{Alpert.10.1137/0524016}. Such polynomial functions can then be dilated and translated to obtain progressively finer representations on a grid:
$$\scaling^n_{jl} = 2^{n/2} \scaling_j(2^n x - l),$$
where $n$ is the current scale and $l$ is the translation index ($l=0 \ldots 2^n-1$). This defines a family of nested vector spaces $\ldots V_{n-1} \subset V_{n} \subset V_{n+1} \ldots $ which is dense in $L^2$. Wavelet functions $\wavelet_{jl}^n$ are obtained as the orthogonal complement of two successive scaling spaces $W_n = V_{n+1}\ominus V_n$. Multidimensional representations are then obtained as standard tensor products, to represent e.g.\ orbitals and the electronic density.

For a practical realisation of an \ac{AE} code, several additional advances were necessary: an adaptive representation of functions\cite{Alpert.10.1006/jcph.2002.7160}, which allows grids to be refined locally, where needed; the non-standard form of operators\cite{Beylkin.10.1016/j.acha.2007.08.001}, which is sparse and enables adaptivity at the stage of operator application; a separated representation of operators, which mitigates the ``curse of dimensionality'' by reducing the prefactor of operator applications from $k^{2d}$ to $k^{d+1}$, where $d$ is the dimensionality of the system\cite{Fann.10.1088/1742-6596/16/1/062}; the integral formulation of the \ac{SCF} equations\cite{Kalos.10.1103/physrev.128.1791,jensen_kinetic_2023} coupled with a \ac{KAIN} method for faster convergence\cite{Harrison.10.1002/jcc.10108}; and specific organisation of the huge amount of data, that describes a set of orbitals in the computational code, to allow efficient parallel processing on HPC systems~\cite{Wind2023}.

\subsection{Core Binding Energy Calculations}

The simplest approach to the calculation of core \acp{BE} is the Koopmans' approach, which uses the negative of the core KS orbital energies as an approximation to the core \acp{BE}. While this has been shown to give good results for relative core \acp{BE} for some systems, it does not, however, take into account final state effects. For systems where final state effects are important, a better approach is $\Delta$SCF, in which an explicit core hole is introduced to a given atom and orbital of interest. The associated BE is then calculated using $\mathrm{BE} = E^{N-1}_{\mathrm{final}} - E^{N}_{\mathrm{initial}}$, where $E^{N-1}_{\mathrm{final}}$ is the energy of the system in its final state, i.e.\ with an explicit core hole, and $E^{N}_{\mathrm{initial}}$ is the initial ground-state energy.

Computing core-ionisation energies brings about the challenge of stabilising the core-hole state: a standard SCF optimisation would not work, because the hole state would likely migrate towards the valence region, resulting in a standard calculation of the ionic system. Several strategies to lock the core-hole state have been devised: 
\begin{itemize}
    \item \textbf{Broken-symmetry guess:} a starting guess for the core-hole calculation is obtained by performing a ground-state calculation with a small additional charge at the core site, as in e.g.\ Ref.~\cite{Kahk2019}. The resulting orbitals are then used as a guess for the core-hole calculation. While this symmetry breaking can help reduce core-hole hopping, no further constraints are applied, so it does not necessarily prevent the collapse of the core state.
    \item \textbf{$Z+1$ technique:} instead of the core-hole calculation, one performs a calculation of a system with an augmented nuclear charge at the core site. This method has better stability, but the results are less accurate that $\Delta$SCF because the system is significantly perturbed. Alternatively, $Z+1$ could be used to generate a better input guess, but like with the broken symmetry guess, if no further constraints are applied, there may still be convergence problems.
    \item \textbf{Pseudopotentials:} all core electrons apart from those associated with the target atom are described via \acp{PSP}. The core-hole state is thus well separated in energy from all other states, preventing core-hole hopping and stabilising convergence.
    \item \textbf{\ac{MOM} and \ac{IMOM}:} the optimisation is constrained by keeping the maximum possible overlap between both the core-hole and other orbitals with their counterparts from either the previous iteration along the optimisation (\ac{MOM}), or the starting guess of the calculation (\ac{IMOM}) 
\end{itemize}

The first two methods are the simplest to adopt as they require minimal effort to modify the implementation of $\Delta$SCF. The last two are more general. The use of \acp{PSP} in combination with \acp{MW} has been demonstrated in a previous study~\cite{Ratcliff2019}. In the present work we combine \acp{MW} with both \ac{MOM} and \ac{IMOM}.

\subsection{\ac{MOM} and \ac{IMOM}}

The goal of a $\Delta$SCF calculation is to optimise a non-Aufbau occupation using the SCF procedure. In most cases, however, the SCF procedure is employed to find the solution for an Aufbau occupation. The challenge posed by core-hole calculations is constituted by the instability of the core-ionised state which could easily collapse, yielding a valence-ionisation state, which has a much lower energy. Additionally, the core-hole could switch site or even delocalise, rendering the results useless when compared to experimental measurements~\cite{Tolbatov2017,Bagus1972}.

To stabilise the core-hole state, \ac{MOM}~\cite{gilbert_self-consistent_2008} attempts to maintain the largest possible overlap between the orbitals at a given iteration in the SCF optimisation and the previous iterate. This is achieved by computing the overlap matrix $\mathbf{O}$ between the two sets of orbitals at successive iterations $k$:
\begin{equation}
    O_{ir}^{(k-1,k)} = \braket{\psi_i^{(k-1)}}{\psi_r^k},
\end{equation}
where the indices $i$ and $r$ denote doubly occupied orbitals and all orbitals, respectively. For each orbital, a weight measure is computed by taking the norm of the corresponding column in the overlap matrix:
\begin{equation}
    p_r = \left(\sum_i^n O_{ir}^2\right)^{1/2}
\end{equation}
The $p_r$ values are then used to assign the occupation in the next iteration, in such a way that the core-hole orbitals stay as close as possible to the previous iterate. This helps prevent both variational collapse and hopping or delocalisation. The \ac{IMOM}~\cite{barca2018} procedure is essentially identical. The only difference is that the reference orbitals to compute the weights $p_r$ are always the initial guess orbitals, rather than being updated at every iteration,
\begin{equation}
    O_{ir}^{(0,k)} = \braket{\psi_i^0}{\psi_r^k}.
\end{equation}
This, in theory, should decrease the likelihood that a variational collapse will occur~\cite{barca2018}.

\subsection{Computational Details}

For both MADNESS and MRChem calculations, ground-state calculations employed localised orbitals, while core-hole calculations employed canonical orbitals. The ground-state orbitals were used as an initial guess for the core-hole calculations for both codes. Calculations were performed using the PBE functional~\cite{Perdew1996}.
Unless otherwise stated, MRChem calculations employed a world precision of 10$^{-5}$ (abbreviated as MW5), and all results are obtained using the \ac{IMOM} protocol, as preliminary calculations showed rare issues with respect to variational collapse using the \ac{MOM} method. MADNESS calculations employed the mixed \ac{AE}/\ac{PSP} protocol, using HGH-GTH \acp{PSP}~\cite{Goedecker1996,Hartwigsen1998}. The MADNESS ground-state calculations used a precision threshold of $10^{-4}$ followed by $10^{-6}$ (polynomial orders $k=6$ and $k=8$ respectively), while core-hole calculations directly used a precision threshold of $10^{-6}$ (wavelet order $k=8$). Both the density and wave function residuals employed convergence criteria of $10^{-3}$. Gaussian basis set calculations were performed using the NWChem code~\cite{Valiev2010}. For each core-hole calculation, an input guess was constructed by performing a ground-state calculation in which a fictitious charge of 0.1$e$ was added to the target atom, following Ref.~\cite{Kahk2019}. Calculations were performed using the def2 basis set family~\cite{def2svp_tzvp_tzvppd}. Spin restricted calculations were performed for all systems, while for glycine and alanine unrestricted calculations were also performed. The molecular structures of ethanol and vinyl fluoride were relaxed using MRChem in the restricted formalism, using localised orbitals with a world precision of 10$^{-6}$. Structures for glycine and alanine were extracted from their respective crystal structures and geometry optimised using BigDFT~\cite{Ratcliff2020} with a grid spacing of 0.5~bohr and coarse (fine) radius multipliers of 5 (7), in all cases using PBE. The atomic structure for 2CzPN was the relaxed structure taken directly from Ref.~\cite{Fernando2022}.

\section{Results}

In the following, we present results for a range of system sizes, from small molecules, where we demonstrate the ability of the \ac{MW} framework to achieve very high precision, through to a large molecule, where we show both the ability to treat large systems and the robustness of the combined \ac{MW} \ac{IMOM} approach, even for systems with a large number of atoms of the same species.

\subsection{Small Molecules}

\begin{figure*}[htb]
\centering
 \begin{subfigure}[t]{1.0\linewidth}
    \centering
    \includegraphics[scale=0.55]{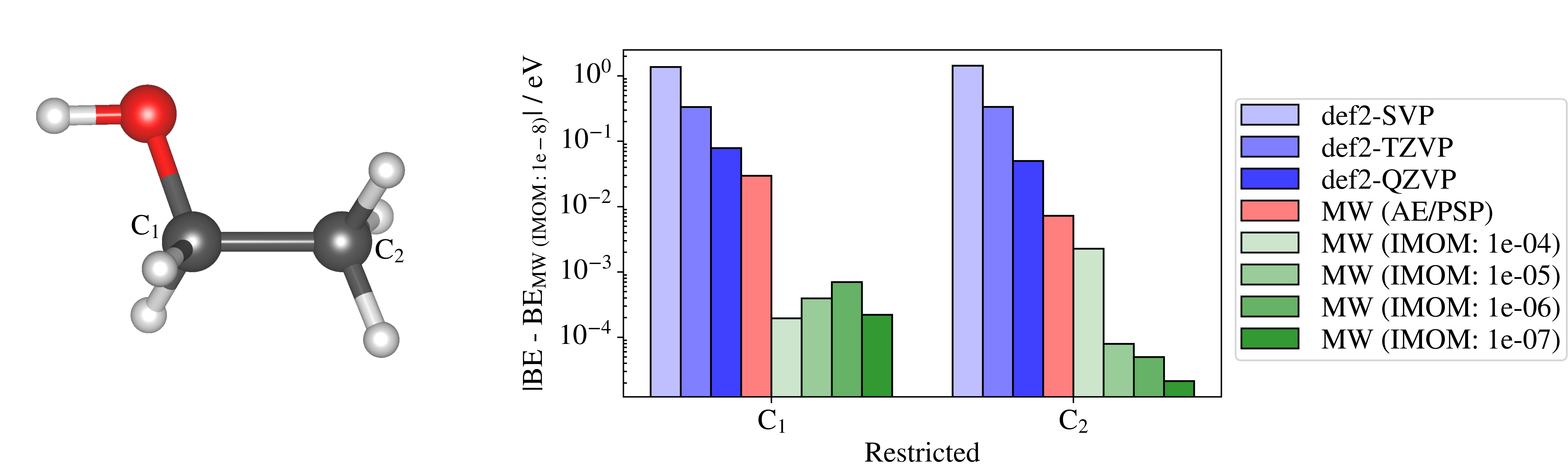}
    \caption{Ethanol}
    \label{fig:eth}
\end{subfigure}
 \begin{subfigure}[t]{1.0\linewidth}
    \centering
    \includegraphics[scale=0.55]{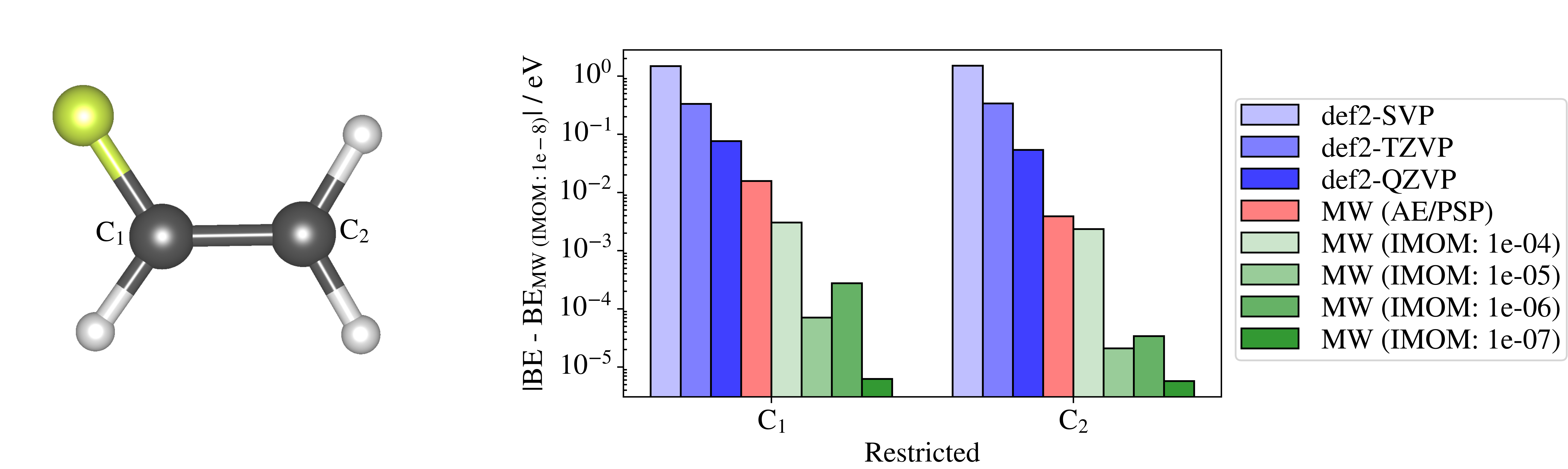}
    \caption{Vinyl fluoride}
    \label{fig:vin}
\end{subfigure}
\caption{Comparison of calculated absolute \acp{BE} for ethanol and vinyl fluoride, for the def2 family of basis sets, \ac{MW} calculations using the \ac{AE}/\ac{PSP} implementation in MADNESS, and the \ac{MW} \ac{IMOM} implementation in MRChem, where the latter has been calculated for a series of different world precisions. Results are given relative to the \ac{MW} \ac{IMOM} implementation in MRChem with a world precision of $10^{-8}$. Shown on the left are the employed atomic structures, labelled with the different atomic environments. See Table~I in the SI for tabulated values.}
\label{fig:small_results}
\end{figure*}

For our first tests, we take ethanol and vinyl fluoride, chosen for being very small molecules, which nonetheless have two distinct C environments. Table~\ref{tab:small_mol} shows the ground-state and core-hole energies for the two molecules, calculated using the \ac{MW} \ac{IMOM} approach for a series of world precisions, as well as using the def2 basis set family. Fig.~\ref{fig:small_results} shows the corresponding \acp{BE}, and also includes results for the \ac{MW} \ac{AE}/\ac{PSP} approach. Compared to the most precise \ac{MW} \ac{IMOM} calculation, the def2 energies differ significantly, by around 5-7~eV for the double zeta basis, decreasing to at most 0.14~eV for the quadruple zeta basis. However, as expected the \acp{BE} benefit from error cancellation, with the difference with respect to the \ac{MW} \ac{IMOM} results decreasing to around 1.5~eV for the double zeta basis and less than 0.08~eV for the quadruple zeta basis. These errors are very similar across C environments in both molecules, so that the error in relative \ac{BE} is less than 0.1~eV even for the double zeta basis compared to the most precise \ac{MW} \ac{IMOM} results (see Table~I and Fig.~1 in the SI). The two different \ac{MW} approaches show excellent agreement, with the \ac{AE}/\ac{PSP} \acp{BE} results differing from the highest precision \ac{IMOM} results by less than 0.03~eV, around the level of thermal broadening effects at room temperature, and well below typical \ac{XPS} resolution. In fact, for absolute \acp{BE}, both \ac{MW} approaches as well as def2-QZVP values all agree to within 0.08~eV, which is less than half the experimental resolution which might be expected for synchrotron \ac{HAXPES} experiments, while the relative \acp{BE} all agree to within 0.03~eV.

\begin{table*}[tb]
\caption{Comparison between total energies for ground state and core hole calculations for ethanol and vinyl fluoride, calculated for the def2 basis set family and a series of world precisions for the \ac{MW} \ac{IMOM} approach. \ac{MW} \ac{AE}/\ac{PSP} energies are omitted since the use of \acp{PSP} precludes a direct comparison. Differences are reported relative to \ac{MW} \ac{IMOM} results using a world precision of 10$^{-8}$. All values are in eV. \label{tab:small_mol}}
\begin{tabular*} {1.0\textwidth}{l @{\extracolsep{\fill}} rrrrrr}
\hline \hline
 & Ground State & $\Delta_{\mathrm{GS}}$ & C$_1$ & $\Delta_{C_1}$ & C$_2$ & $\Delta_{C_2}$\\
\cline{1-1}\cline{2-2}\cline{3-3}\cline{4-4}\cline{5-5}\cline{6-6}\cline{7-7}\\[-2ex]

\textbf{Ethanol}\\

def2-SVP	 & -4210.34083260 & 5.31245782	&-3910.69896692	& 6.68120103 & -3912.12437184 & 6.74606771\\
def2-TZVP	& -4215.24887563 & 0.40441479 &-3916.64107365	& 0.73909431 & -3918.12982277 & 0.74061679\\
def2-QZVP& 	-4215.59047745	& 0.06281297 & -3917.23880636	& 0.14136159 & -3918.75773216 & 0.11270739\\	

\ac{MW} (\ac{IMOM}) $10^{-4}$ & -4215.64993550 & 0.00335492 & -3917.37700801 & 0.00315995 & -3918.86482016 & 0.00561940\\
\ac{MW} (\ac{IMOM}) $10^{-5}$ & -4215.65327307 & 0.00001735 & -3917.38054491 & -0.00037696 & -3918.87034277 & 0.00009679\\
\ac{MW} (\ac{IMOM}) $10^{-6}$ & -4215.65328843 & 0.00000200 & -3917.38086603 & -0.00069807 & -3918.87038779 & 0.00005176\\
\ac{MW} (\ac{IMOM}) $10^{-7}$ & -4215.65329010 & 0.00000033 & -3917.38038850 & -0.00022055 & -3918.87041786 & 0.00002169\\
\ac{MW} (\ac{IMOM}) $10^{-8}$ & -4215.65329042 & / & -3917.38016795 & / & -3918.87043956 & /\\

\cline{1-1}\cline{2-2}\cline{3-3}\cline{4-4}\cline{5-5}\cline{6-6}\cline{7-7}\\[-2ex]

\textbf{Vinyl fluoride}\\
def2-SVP	&-4829.83770206 & 6.01642799 &	-4529.01676646 & 7.49937027	&-4531.33997522 & 7.52241718\\
def2-TZVP	&-4835.50767314	& 0.34645691 & -4535.83728200 & 0.67885473 & -4538.17714577 & 0.68524663\\
def2-QZVP&	-4835.79935913 & 0.05477092 &	-4536.38514167& 0.13099506 & -4538.75373277 & 0.10865963\\

\ac{MW} (\ac{IMOM}) $10^{-4}$ & -4835.84334010 & 0.01078995 & -4536.50838962 & 0.00774711 & -4538.85394715 & 0.00844525\\
\ac{MW} (\ac{IMOM}) $10^{-5}$ & -4835.85395294 & 0.00017711 & -4536.51603052 & 0.00010621 & -4538.86219431 & 0.00019809\\
\ac{MW} (\ac{IMOM}) $10^{-6}$ & -4835.85412789 & 0.00000216 & -4536.51585879 & 0.00027794 & -4538.86235630 & 0.00003611\\
\ac{MW} (\ac{IMOM}) $10^{-7}$ & -4835.85412898 & 0.00000107 & -4536.51612942 & 0.00000732 &  -4538.86238559 & 0.00000682\\
\ac{MW} (\ac{IMOM}) $10^{-8}$ & -4835.85413005 & / & -4536.51613673 & / & -4538.86239240 & /\\		

 \hline \hline
\end{tabular*}
\end{table*}

Considering now the convergence with respect to world precision for the \ac{MW} \ac{IMOM} results, the \acp{BE} do not show the expected systematic convergence behaviour. Looking at the total energies, the ground-state energies converge smoothly, while the core-hole energies are less systematic. One possibility is that core-hole calculations are not strictly bound from below like the ground state and this might cause instabilities. Although \ac{IMOM} prevents the full collapse of the core-hole state it might still not be able to avoid such small artifacts. This is e.g.\ particularly evident for the C1-hole state (MW5, MW6 and MW7 precision) of ethanol, which display lower total energies than MW8 precision. Some deviations might be expected where e.g.\ a lower precision calculation might by chance more than satisfy the precision requirements, however it is not clear why the core-hole calculations are so susceptible to showing non-systematic convergence, while the ground-state calculations are not. Additional calculations were performed using canonical ground-state orbitals, and similar behaviour was observed. 

This behaviour affects the error on the BE, which is now a difference between two absolute energy errors. As a result the lack of systematicity is amplified leading to a somehow erratic convergence to the MW8 result.
Nonetheless, these differences are very small, with the total energies differing by around 0.01~eV even for the least precise calculations ($10^{-4}$ world precision), decreasing to the order of $10^{-4}$~eV for the next smallest precision ($10^{-5}$). These differences are smaller than any of the differences with respect to the other approaches, and more than sufficient for practical calculations. Therefore, a world precision of $10^{-5}$ is used for all subsequent calculations.

\subsection{Amino Acids}

\begin{figure*}[htb]
\centering
 \begin{subfigure}[t]{1.0\linewidth}
    \centering
    \includegraphics[scale=0.55]{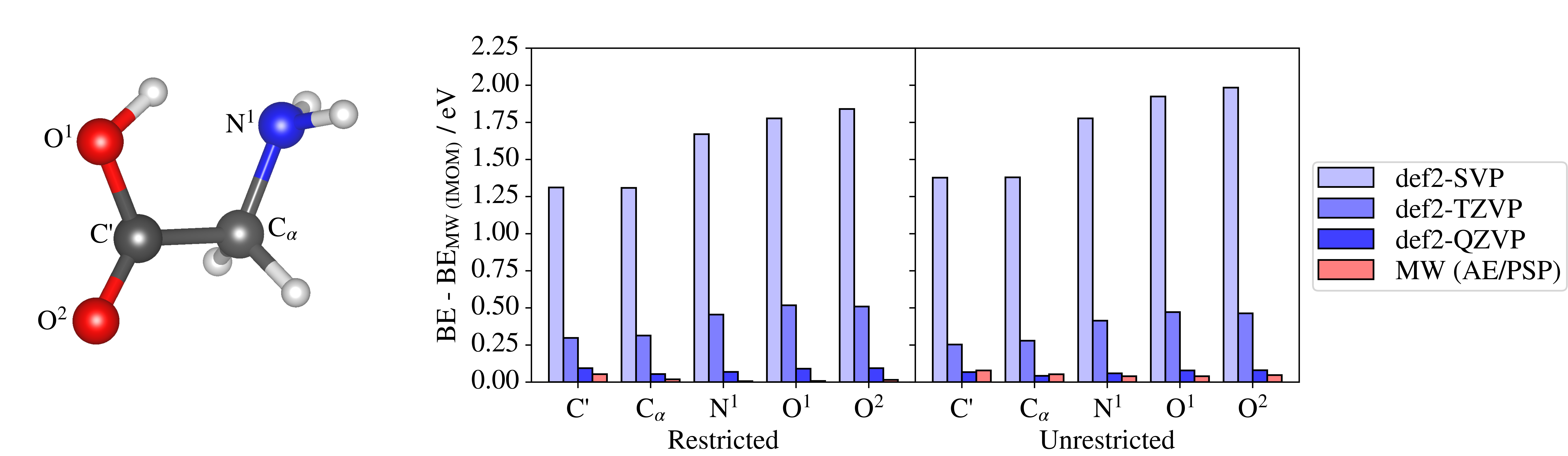}
    \caption{Gly}
    \label{fig:gly}
\end{subfigure}
 \begin{subfigure}[t]{1.0\linewidth}
    \centering
    \includegraphics[scale=0.55]{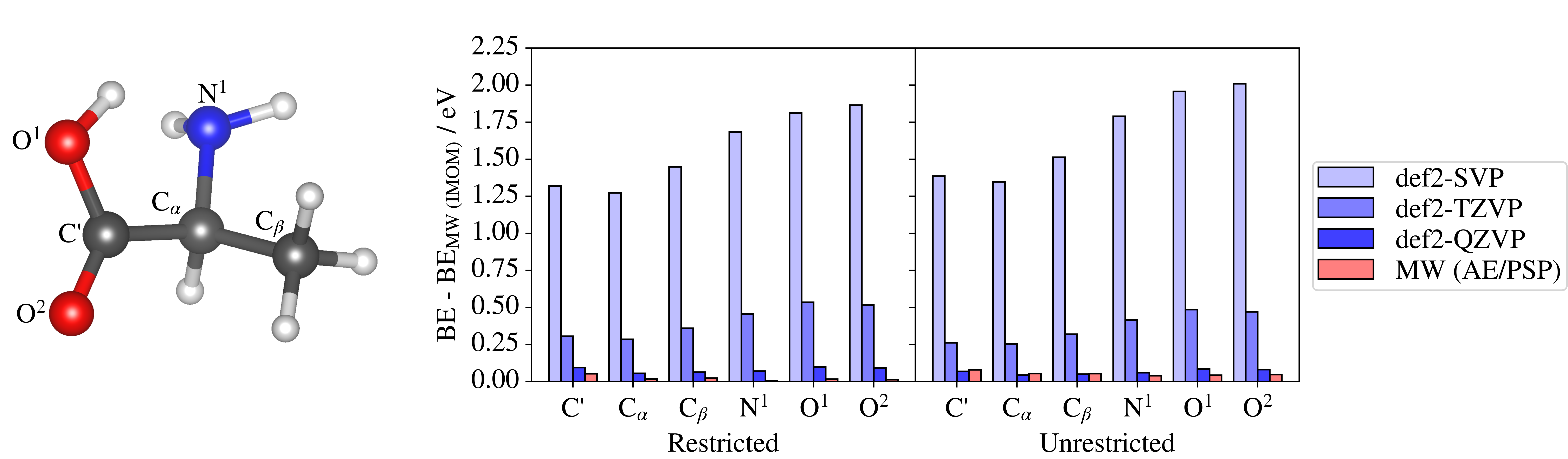}
    \caption{Ala}
    \label{fig:ala}
\end{subfigure}
\caption{Comparison of calculated absolute \acp{BE} for the amino acids glycine and alanine for both restricted and unrestricted calculations, for the def2 family of basis sets, and the two different \ac{MW} implementations. Results are given relative to the \ac{MW} \ac{IMOM} implementation in MRChem. Shown on the left are the atomic structures of the employed conformers, labelled with the different atomic environments.}
\label{fig:aa_results}
\end{figure*}

As our second example, we take the amino acids glycine (Gly) and alanine (Ala), which have been well characterised from both a theoretical and experimental perspective, including previous calculations using the \ac{AE}/\ac{PSP} approach implemented in Ref.~\cite{Pi2020}. In this work, we take a single conformer, and perform calculations in both restricted and unrestricted formalisms, again comparing \ac{MW} \ac{IMOM}, \ac{MW} \ac{AE}/\ac{PSP} and def2 basis set calculations. Results are depicted in Fig.~\ref{fig:aa_results}, with the absolute \acp{BE} given in Table~II in the SI.

Similar to the small molecules, the double zeta \acp{BE} differ significantly from the \ac{MW} \ac{IMOM} values, by 1-2~eV. The error is already much smaller for the triple zeta basis, with an average difference of 0.4~eV across all calculations, further reducing to 0.07~eV for the quadruple zeta basis. The difference between \ac{MW} approaches is again small, with an average difference of 0.04~eV between the two. The relative \acp{BE} again benefit from error cancellation. There are some variations between different core states, but in the worst case, the relative \acp{BE} of the triple zeta differ by 0.06~eV with respect to the \ac{MW} \ac{IMOM} results, while the quadruple zeta and \ac{AE}/\ac{PSP} results differ by even less (see Fig.~2 in the SI).

Finally, using a restricted approach results in a large shift in the absolute \acp{BE}, of around 7-9~eV depending on the chemical species. However, the effect on the relative \acp{BE} is much smaller, with the mean absolute deviation between restricted and unrestricted relative \acp{BE} being less than 0.01~eV for all basis sets. Thus, the error in relative \ac{BE} introduced by using a spin restricted formalism is lower than the basis set effects.

\subsection{2CzPN}

We finish with the example of 2CzPN, a prototypical thermally activated delayed fluorescence (TADF)-based organic light emitting diode (OLED) emitter. 2CzPN has previously been investigated with MADNESS~\cite{Fernando2022}, where it was shown that theory is essential for interpreting the experimental spectra, with three distinct peaks in the C~1$s$ spectra coming from seven underlying chemical environments. Being a larger molecule (54 atoms/119 doubly-occupied orbitals), it is also a good test of both the ability of MRChem to treat large systems, and of the effectiveness of the \ac{IMOM} approach in the case when there are a large number of C atoms, in some cases with the same local chemical environment, and thus a high likelihood of core-hole hopping occurring.

\begin{figure*}[htb]
 \centering
 \includegraphics[scale=0.55]{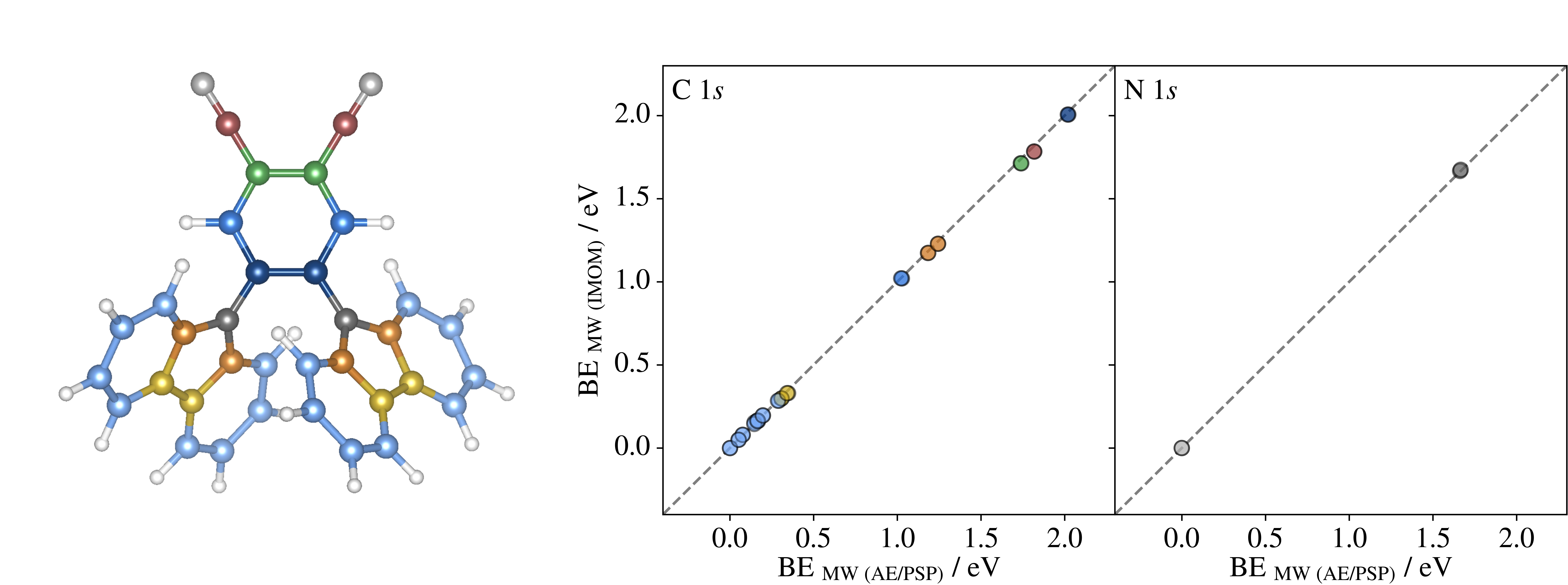}
\caption{Comparison between relative \acp{BE} for the relaxed 2CzPN molecule, for the \ac{MW} \ac{IMOM} implementation in MRChem and the mixed \ac{AE}/\ac{PSP} \ac{MW} MADNESS results from ref.~\cite{Fernando2022}. Shown on the left is the atomic structure of 2CzPN, where the atoms have been coloured according to their different local chemical environments, where the two N atoms are in shades of grey, and the C atoms are in a range of colours.}
\label{fig:2czpn}
\end{figure*}

All \ac{MW} \ac{IMOM} calculations successfully converged, demonstrating the ability of \ac{IMOM} to handle C atoms with multiple similar local chemical environments. Shown in Fig.~\ref{fig:2czpn} are the core \acp{BE} for the relaxed 2CzPN structure, where the \ac{MW} \ac{IMOM} results are compared with the mixed \ac{AE}/\ac{PSP} \ac{MW} results from previous work~\cite{Fernando2022}. As can be seen, the relative \acp{BE} for each approach are in excellent agreement, for both C and N atoms. Furthermore, the absolute \acp{BE} have a mean absolute deviation between the two approaches of 0.02~eV. As discussed above, this is of  the order of thermal broadening effects at room temperature and well below experimental resolution. In short, the \ac{MW} \ac{IMOM} MRChem approach proves to be robust and accurate for core \acp{BE} calculations of large molecules.

\section{Conclusions}

In this work, we have combined the \acl{MOM} and \acl{IMOM} with a Multiwavelet approach to describe the electronic structure of a molecule.
\ac{MOM} and \ac{IMOM} are designed to stabilise the convergence of core-hole state calculations, by ensuring the hole state is kept as close as possible to the original electronic state. Multiwavelets, on the other hand, provide a robust and systematic approach to the \ac{CBS} limit, overcoming limitations of \ac{AO} bases. This is especially critical for core properties, which negate one of the basic assumptions behind using \acp{AO}: error cancellation, which generally mitigates an inadequate description of core states, but it can be exploited when electronic structure changes are essentially affecting the valence electrons.
This combination has proven very successful: we are able to reproduce results obtained using an approach which combines \acp{MW} with the use of pseudopotentials for all but one core-excited atom (for which all electrons are treated explicitly) and to a precision which is at least one order of magnitude better than the resolution of \ac{XPS} experiments. We show that such a precision is achieved consistently, albeit showing a less systematic trend with respect to increasing precision than expected. We are also able to treat a large molecule with many chemically similar atoms, which is a notorious challenge for this kind of calculation.
Between \ac{MOM} and \ac{IMOM}, the latter has proved slightly superior in terms of reliability: although both are implemented, all results featured in the paper are obtained with \ac{IMOM}, as the \ac{MOM} procedure can show a variational collapse in rare cases. We have designed a robust protocol which combines the use of localised orbitals in the ground state, which allows one to select a core orbital located at a specific atom, with canonical orbitals for the excited state, which enables convergence to the requested core-hole state by decoupling \ac{SCF} equations during the optimisation. Together with \ac{IMOM}, this effectively prevented the collapse of the core hole for all cases investigated.
Work is ongoing to extend the current work to include \ac{HF} exchange. Thereafter we would like to apply the method to a range of larger molecules, including for example extending previous work on exploiting the local nature of core \ac{XPS} for probing the effects of disorder in 2CzPN to other TADF emitters. The complexity of the measured spectra for such molecules necessitates the use of theory to enable the interpretation of experimental results, while the large size of such molecules requires a robust and efficient computational approach, which our current approach provides.

\section*{Data availability}

The data supporting this article have been included in the ESI.

\section*{Conflicts of interest}

There are no conflicts of interest to declare.

\section*{Acknowledgements}

We are grateful for computational support from the UK national high performance computing service, ARCHER2, for which access was obtained via the UKCP consortium and funded by EPSRC grant ref EP/X035891/1. LER thanks Anna Regoutz for useful discussions.
The financial support from the Research Council of Norway through its Centres of Excellence scheme (Hylleraas centre, 262695) and the FRIPRO scheme (ReMRChem, 324590) is acknowledged. The support from the Norwegian Metacenter for Computational Science (NOTUR) infrastructure under the nn14654k grant of computer time is also acknowledged.

%\balance

%%%REFERENCES%%%
\bibliography{lit, luca} 
\bibliographystyle{rsc} %the RSC's .bst file

\end{document}

%
% This is a random collection of acronyms
% Acronyms are ordered alphabetically by abbreviation
% Usage
% 1. add the following lines to your main tex document
% In the preamble: \usepackage{acronym}
% In the document: \input{acronyms}
%
% 2. RTFM of the acronym package!
%
% 3. Add/modify/remove the listes acronyms according to your needs
%
\begin{acronym}
\acro{AO}{atomic orbital}
\acro{AE}{all electron}
\acro{API}{Application Programmer Interface}
\acro{AUS}{Advanced User Support}
\acro{BE}{Binding Energy}
\acrodefplural{BE}[BEs]{Binding Energies}
\acro{BEM}{Boundary Element Method}
\acro{BO}{Born-Oppenheimer}  
\acro{CBS}{complete basis set}
\acro{CC}{Coupled Cluster}
\acro{CTCC}{Centre for Theoretical and Computational Chemistry}
\acro{CoE}{Centre of Excellence}
\acro{DC}{dielectric continuum}  
\acro{DD}{domain decomposition}
\acro{DFT}{density functional theory}  
\acro{DKH}{Douglas-Kroll-Hess}
\acro{EFP}{effective fragment potential}
\acro{ECP}{effective core potential}
\acro{EU}{European Union}
\acro{FMM}{fast multipole method}
\acro{FCI}{Full Configuration Interaction}
\acro{GGA}{generalized gradient approximation}
\acro{GPE}{Generalized Poisson Equation}
\acro{GTO}{Gaussian Type Orbital}
\acro{HF}{Hartree-Fock}  
\acro{HPC}{high-performance computing}
\acro{Hylleraas}[HC]{Hylleraas Centre for Quantum Molecular Sciences}
\acro{IEF}{Integral Equation Formalism}
\acro{IEFPCM}{Integral Equation Formalism \ac{PCM}}
\acro{IGLO}{individual gauge for localized orbitals}
\acro{IMOM}{Initial Maximum Overlap Method}
\acro{KB}{kinetic balance}
\acro{KS}{Kohn-Sham}
\acro{LAO}{London atomic orbital}
\acro{LAPW}{linearized augmented plane wave}
\acro{LDA}{local density approximation}
\acro{MAD}{mean absolute deviation}
\acro{maxAD}{maximum absolute deviation}
\acro{MM}{molecular mechanics}  
\acro{MCSCF}{multiconfiguration self consistent field}
\acro{MPA}{multiphoton absorption}
\acro{MRA}{multiresolution analysis}
\acro{MSDD}{Minnesota Solvent Descriptor Database}
\acro{MOM}{Maximum Overlap Method}
\acro{MW}{multiwavelet}
\acro{NAO}{numerical atomic orbital}
\acro{NeIC}{nordic e-infrastructure collaboration}
\acro{KAIN}{Krylov-accelerated inexact Newton}
\acro{NMR}{nuclear magnetic resonance}
\acro{NP}{nanoparticle}  
\acro{OLED}{organic light emitting diode}
\acro{PAW}{projector augmented wave}
\acro{PBC}{Periodic Boundary Condition}
\acro{PCM}{polarizable continuum model}
\acro{PSP}{pseudopotential}
\acro{PW}{plane wave}
\acro{QC}{quantum chemistry}  
\acro{QM/MM}{quantum mechanics/molecular mechanics}  
\acro{QM}{quantum mechanics}  
\acro{RCN}{Research Council of Norway}
\acro{RDM}{1-body reduced density matrix}
\acro{RMSD}{root mean square deviation}
\acro{RKB}{restricted kinetic balance}
\acro{SC}{semiconductor}
\acro{SCF}{self-consistent field}
\acro{SCRF}{self-consistent reaction field}
\acro{STSM}{short-term scientific mission}
\acro{SAPT}{symmetry-adapted perturbation theory}
\acro{SERS}{surface-enhanced raman scattering}
\acro{STO}{Slater-Type Orbital}
\acro{WPREL}[WP1]{Work Package 1}
\acro{WPROP}[WP2]{Work Package 2}
\acro{WPAPP}[WP3]{Work Package 3}
\acro{WP}{Work Package}  
\acro{X2C}{exact two-component}
\acro{ZORA}{zero-order relativistic approximation}
\acro{HAXPES}{hard X-ray photoelectron spectroscopy}
\acro{XPS}{X-ray photoelectron spectroscopy}
\end{acronym}

\title{{\LARGE Supplementary Information}\\Combining the Maximum Overlap Method with Multiwavelets for Core-Ionisation Energy Calculations}

\author{Niklas G\"{o}llmann}\affiliation{Theoretische Organische Chemie, Organisch-Chemisches Institut and Center for Multiscale Theory and Computation, Universit\"{a}t M\"{u}nster, Corrensstra{\ss}e 36, 48149 M\"{u}nster, Germany}\affiliation{Hylleraas Centre, UiT The Arctic University of Norway, N-9037 Tromsø, Norway}
\author{Matthew~R.~Ludwig}\affiliation{Centre for Computational Chemistry, School of Chemistry, University of Bristol, Bristol BS8 1TS, United Kingdom}
\author{Peter Wind}\affiliation{Hylleraas Centre, UiT The Arctic University of Norway, N-9037 Tromsø, Norway}
\author{Laura~E.~Ratcliff}\email{laura.ratcliff@bristol.ac.uk}\affiliation{Centre for Computational Chemistry, School of Chemistry, University of Bristol, Bristol BS8 1TS, United Kingdom}\affiliation{Hylleraas Centre, UiT The Arctic University of Norway, N-9037 Tromsø, Norway}
\author{Luca Frediani}\affiliation{Hylleraas Centre, UiT The Arctic University of Norway, N-9037 Tromsø, Norway}

\date{\today}

\maketitle

 \begin{figure*}[htb]
\centering
 \begin{subfigure}[t]{1.0\linewidth}
    \centering
    \includegraphics[scale=0.55]{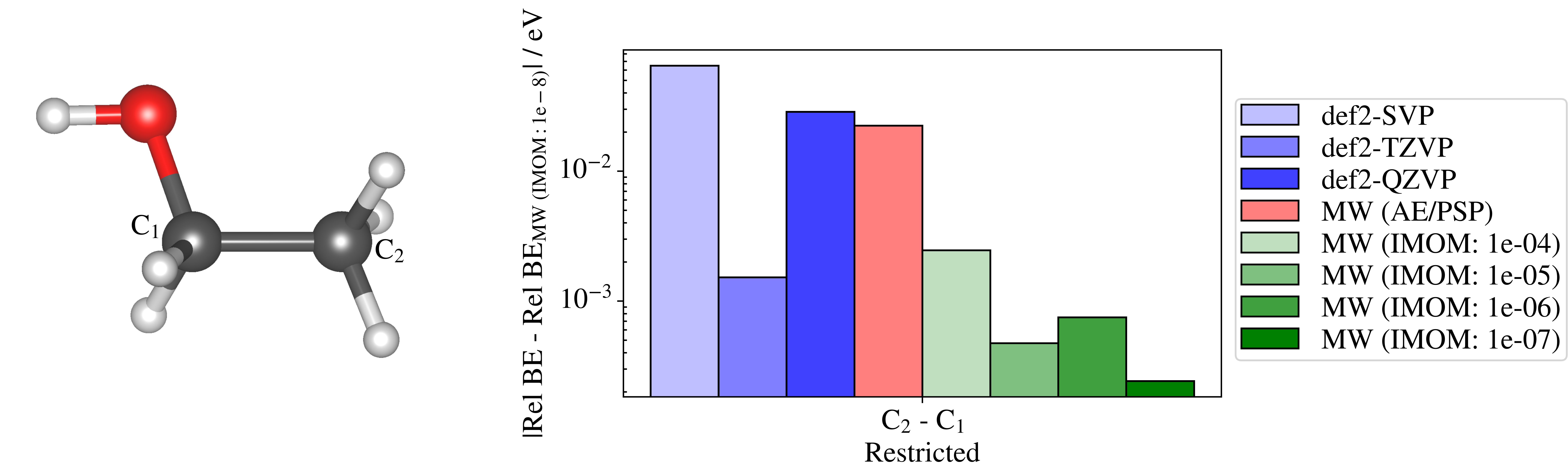}
    \caption{Ethanol}
    \label{fig:gly}
\end{subfigure}
 \begin{subfigure}[t]{1.0\linewidth}
    \centering
    \includegraphics[scale=0.55]{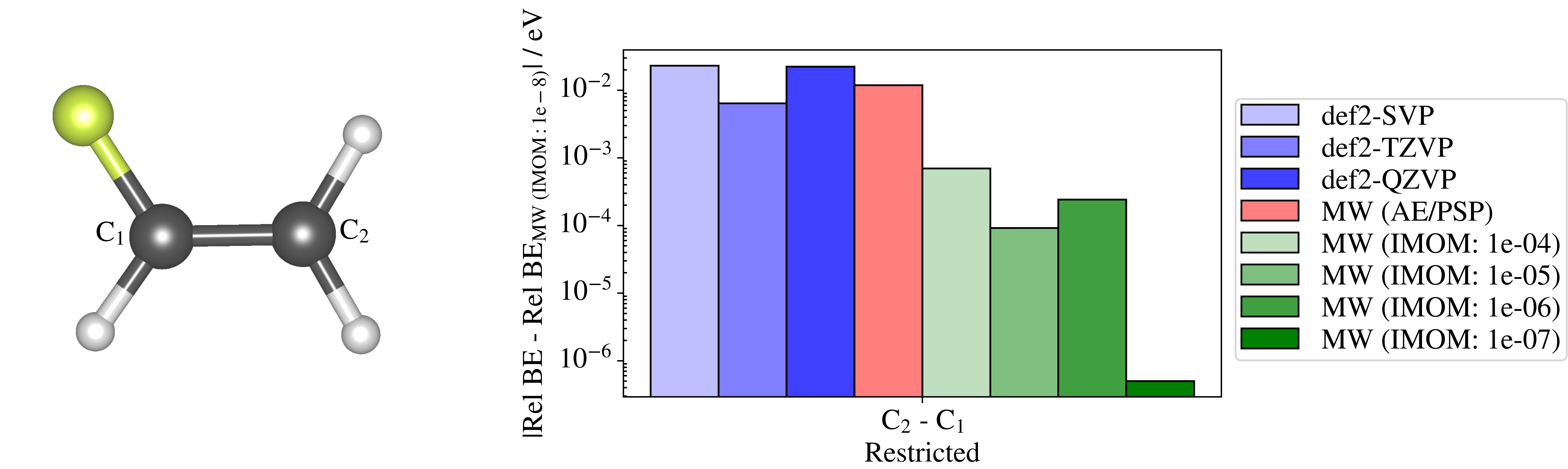}
    \caption{Vinyl fluoride}
    \label{fig:ala}
\end{subfigure}
\caption{Relative \acp{BE} for ethanol and vinyl fluoride, for the def2 family of basis sets, \ac{MW} calculations using the \ac{AE}/\ac{PSP} implementation in MADNESS, and the \ac{MW} \ac{IMOM} implementation in MRChem, where the latter has been calculated for a series of different world precisions. Results are given relative to the \ac{MW} \ac{IMOM} implementation in MRChem with a world precision of $10^{-8}$. Shown on the left are the employed atomic structures, labelled with the different atomic environments. See Table~\ref{tab:small_mol_si} for tabulated values.}
\label{fig:small_results_si}
\end{figure*} 

\begin{table*}
\caption{Absolute \acp{BE} for ethanol and vinyl fluoride, as well as \ac{BE} differences. Results are given for the def2 basis set family, the \ac{MW} \ac{AE}/\ac{PSP} approach and a series of world precisions for the \ac{MW} \ac{IMOM} approach.  All values are in eV. \label{tab:small_mol_si}}
\begin{tabular*} {1.0\textwidth}{l @{\extracolsep{\fill}} rrr}
\hline \hline
 & C$_1$ & C$_2$ & Difference \\
\cline{1-1}\cline{2-2}\cline{3-3}\cline{4-4}\\[-2ex]

\textbf{Ethanol}\\

def2-SVP	 &299.641866	& 298.216461	&1.425405\\
def2-TZVP	& 298.607802	& 297.119053	&1.488749\\
def2-QZVP& 	298.351671	& 296.832745 & 1.518926\\

\ac{MW} (\ac{AE}/\ac{PSP}) & 298.302768&	296.790094	& 1.512674\\ 

\ac{MW} (\ac{IMOM}) $10^{-4}$ & 298.272927	&296.785115&	1.487812
\\
\ac{MW} (\ac{IMOM}) $10^{-5}$ & 298.272728	&296.782930&	1.489798
\\
\ac{MW} (\ac{IMOM}) $10^{-6}$ &298.272422	&296.782901	&1.489522
\\
\ac{MW} (\ac{IMOM}) $10^{-7}$ & 298.272902	&296.782872	&1.490029
\\
\ac{MW} (\ac{IMOM}) $10^{-8}$ & 298.273122	&296.782851&	1.490272
\\

\cline{1-1}\cline{2-2}\cline{3-3}\cline{4-4}\\[-2ex]

\textbf{Vinyl fluoride}\\
def2-SVP	&300.820936 &	298.497727	&2.323209
\\
def2-TZVP	&299.670391&	297.330527	&2.339864
\\
def2-QZVP&	299.414217	&297.045626	&2.368591
\\
\ac{MW} (\ac{AE}/\ac{PSP}) & 299.353748	&296.995630	&2.358118\\

\ac{MW} (\ac{IMOM}) $10^{-4}$ &299.334950&	296.989393&	2.345558
\\
\ac{MW} (\ac{IMOM}) $10^{-5}$ &299.337922&	296.991759	&2.346164
\\
\ac{MW} (\ac{IMOM}) $10^{-6}$ & 299.338269&	296.991772	&2.346498
\\
\ac{MW} (\ac{IMOM}) $10^{-7}$ & 299.338000	&296.991743&	2.346256
\\
\ac{MW} (\ac{IMOM}) $10^{-8}$ & 299.337993	&296.991738&	2.346256
\\

 \hline \hline
\end{tabular*}
\end{table*}

 \begin{figure*}[htb]
\centering
 \begin{subfigure}[t]{1.0\linewidth}
    \centering
    \includegraphics[scale=0.55]{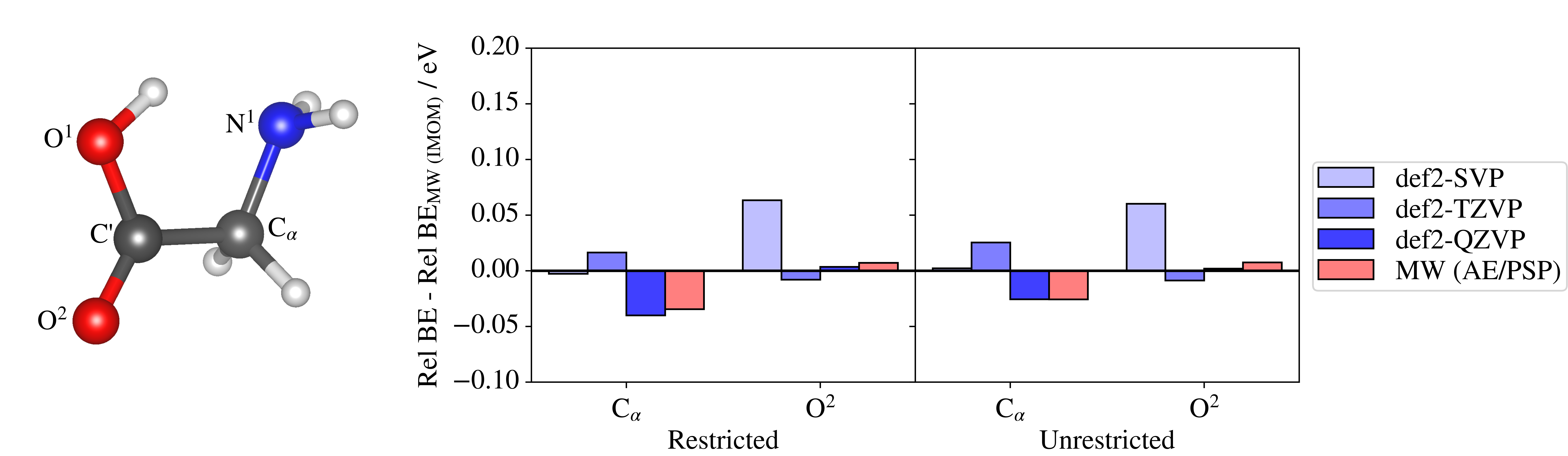}
    \caption{Gly}
    \label{fig:gly}
\end{subfigure}
 \begin{subfigure}[t]{1.0\linewidth}
    \centering
    \includegraphics[scale=0.55]{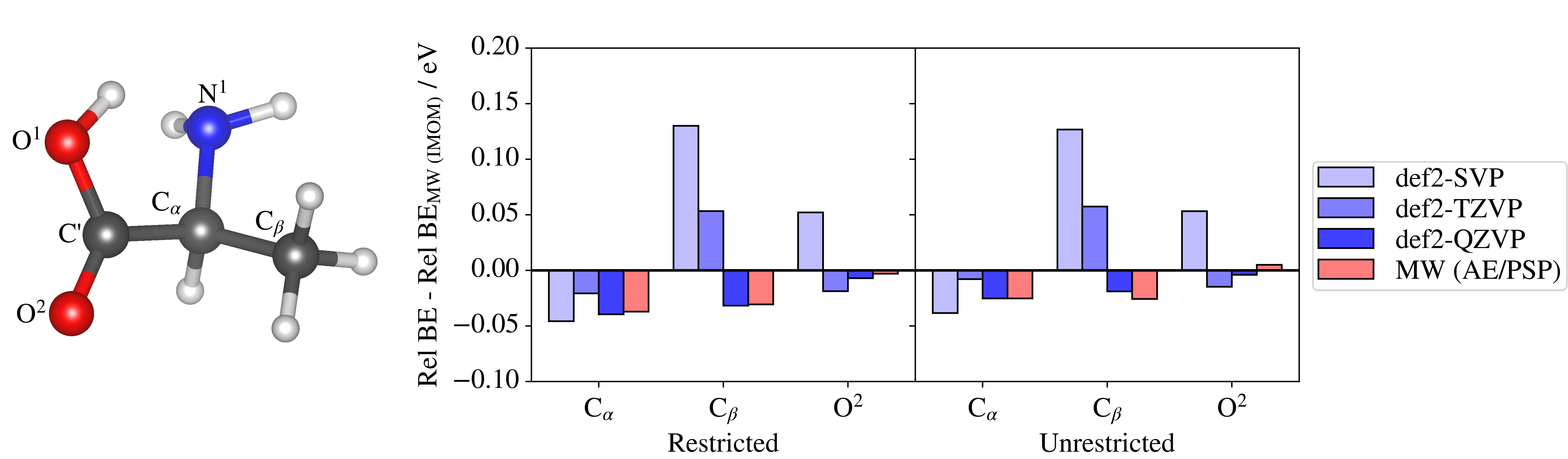}
    \caption{Ala}
    \label{fig:ala}
\end{subfigure}
\caption{Relative \acp{BE} for the amino acids glycine and alanine for both restricted and unrestricted calculations, for the def2 family of basis sets, and the two different \ac{MW} implementations. C \acp{BE} are given relative to C' while O \acp{BE} are given relative to O$^1$. All results are given relative to the \ac{MW} \ac{IMOM} implementation in MRChem. Shown on the left are the atomic structures of the employed conformers, labelled with the different atomic environments. See Table~\ref{tab:aa_mol_si} for tabulated \acp{BE}.}
\label{fig:aa_results_si}
\end{figure*}

\begin{table*}
\caption{Calculated core \acp{BE} for the amino acids glycine and alanine. Restricted and unrestricted results are given for the \ac{MW} formalism combined with \ac{IMOM} in MRChem, the \ac{MW} formalism combined with the \ac{AE}/\ac{PSP} formalism in MADNESS, and the def2 basis set family in NWChem. \label{tab:aa_mol_si}}
\begin{tabular*} {1.0\textwidth}{l @{\extracolsep{\fill}} rrrrr}
\hline \hline
 & \ac{MW} (\ac{IMOM}) & \ac{MW} (\ac{AE}/\ac{PSP}) & def2-SVP & def2-TZVP & def2-QZVP \\
\cline{1-1}\cline{2-2}\cline{3-3}\cline{4-4}\cline{5-5} \cline{6-6}\\[-2.5ex]

\textbf{Restricted}\\

\textbf{Gly}\\
C' & 300.393125 & 300.446573 & 301.704520 & 300.690755 & 300.487082\\
C$_\alpha$ & 298.473433 & 298.492359 & 299.782111 & 298.787486 & 298.527342\\
N$^1$ & 413.297428 & 413.303946 & 414.968204 & 413.752325 & 413.366353\\
O$^1$ & 546.966640 & 546.974559 & 548.743397 & 547.484024 & 547.056733\\
O$^2$ & 545.343520 & 545.358534 & 547.183611 & 545.852887 & 545.437166\\
\\

\textbf{Ala}\\
C' & 300.182194 & 300.235265 & 301.501556 & 300.487685 & 300.276533\\
C$_\alpha$ & 298.331945 & 298.347913 & 299.605617 & 298.616736 & 298.386792\\
C$_\beta$ & 297.286080 & 297.308566 & 298.735451 & 297.644877 & 297.348702\\
N$^1$ & 413.008375 & 413.015518 & 414.691391 & 413.464175 & 413.077908\\
O$^1$ & 546.819229 & 546.834812 & 548.631460 & 547.353350 & 546.917982\\
O$^2$ & 545.221391 & 545.233976 & 547.085744 & 545.736807 & 545.313132\\
\\

\cline{1-1}\cline{2-2}\cline{3-3}\cline{4-4}\cline{5-5} \cline{6-6}\\[-2.5ex]

\textbf{Unrestricted}\\

\textbf{Gly}\\
C' & 293.379773 & 293.458123 & 294.757253 & 293.633292 & 293.447485\\
C$_\alpha$ & 291.458030 & 291.510726 & 292.837706 & 291.736918 & 291.500174\\
N$^1$ & 405.196269 & 405.236047 & 406.973117 & 405.610341 & 405.255558\\
O$^1$ & 537.818944 & 537.858625 & 539.743112 & 538.290690 & 537.897337\\
O$^2$ & 536.195769 & 536.243007 & 538.180115 & 536.658801 & 536.276099\\
\\

\textbf{Ala}\\
C' & 293.169480 & 293.248933 & 294.555586 & 293.431025 & 293.237792\\
C$_\alpha$ & 291.311977 & 291.366203 & 292.659717 & 291.565613 & 291.355015\\
C$_\beta$ & 290.279409 & 290.333125 & 291.792274 & 290.598352 & 290.328774\\
N$^1$ & 404.908816 & 404.948379 & 406.698122 & 405.323974 & 404.968724\\
O$^1$ & 537.674243 & 537.716739 & 539.631212 & 538.160122 & 537.758733\\
O$^2$ & 536.071585 & 536.119090 & 538.081832 & 536.542688 & 536.152070\\

 \hline \hline
\end{tabular*}
\end{table*}

 \begin{figure*}[htb]
\centering
    \includegraphics[scale=0.65]{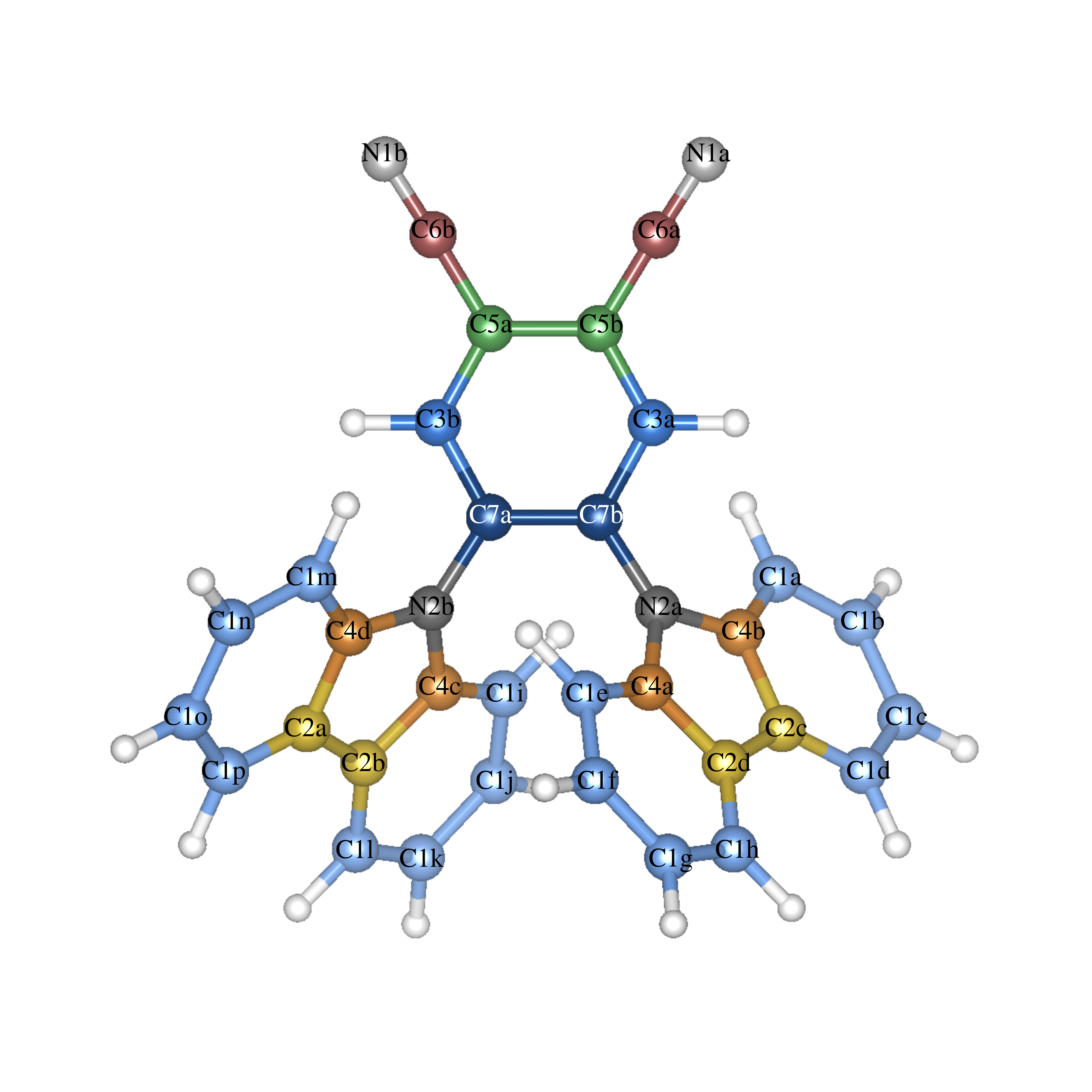}
\caption{Atom labelling for 2CzPN.}
\label{fig:2czpn_si}
\end{figure*}

\begin{table*}
\caption{Core \acp{BE} for 2CzPN, calculated using both \ac{MW} approaches. All values are given in eV, while the atom labelling is depicted in Fig.~\ref{fig:2czpn_si}. \label{tab:2czpn_si}}
\begin{tabular*} {1.0\textwidth}{l @{\extracolsep{\fill}} rrrrr}
\hline \hline
  & \ac{MW} (\ac{AE}/\ac{PSP}) & \ac{MW} (\ac{IMOM})\\
\cline{1-1}\cline{2-2}\cline{3-3}\\[-2.5ex]
C1a & 296.137530 & 296.116675 \\
C1b & 296.006727 & 295.993991 \\
C1c & 295.897237 & 295.881515 \\
C1d & 296.043533 & 296.028364 \\
C1e & 296.014609 & 295.994500 \\
C1f & 295.922288 & 295.911398 \\
C1g & 295.846540 & 295.831671 \\
C1h & 295.992592 & 295.985925 \\
C1i & 296.011576 & 295.998264 \\
C1j & 295.922404 & 295.911213 \\
C1k & 295.845936 & 295.831350 \\
C1l & 295.991520 & 295.975904 \\
C1m & 296.132328 & 296.116679 \\
C1n & 296.010861 & 295.994580 \\
C1o & 295.898230 & 295.881605 \\
C1p & 296.042501 & 296.027664 \\

C2a & 296.187557 & 296.161922 \\
C2b & 296.154696 & 296.130305 \\
C2c & 296.194186 & 296.162034 \\
C2d & 296.153710 & 296.129887 \\

C3a & 296.874169 & 296.851927 \\
C3b & 296.869010 & 296.852594 \\

C4a & 297.029044 & 297.004898 \\
C4b & 297.090294 & 297.060365 \\
C4c & 297.031018 & 297.005966 \\
C4d & 297.088987 & 297.060524 \\

C5a & 297.586669 & 297.544598 \\
C5b & 297.583503 & 297.544740 \\

C6a & 297.664987 & 297.615004 \\
C6b & 297.663048 & 297.616121 \\

C7a & 297.868197 & 297.838072 \\
C7b & 297.864101 & 297.837994 \\

N1a & 411.295818 & 411.277388 \\
N1b & 411.294868 & 411.278290 \\

N2a & 412.957678 & 412.944421 \\
N2b & 412.961914 & 412.952033 \\

 \hline \hline
\end{tabular*}
\end{table*}

%\bibliographystyle{apsrev4-1}
%\bibliography{refs}